\documentclass[useAMS,usenatbib]{mnras}
\usepackage{graphicx}
\usepackage{txfonts}
\usepackage[dvipsnames]{color}
\usepackage{ulem}
\usepackage{ragged2e}
\def\aap{A\&A}
\def\apj{ApJ}
\def\apjl{ApJL}
\def\apjs{ApJS}
\def\mnras{MNRAS}

\def\nat{Nature}

\usepackage{soul}
\usepackage[dvipsnames]{xcolor}
\usepackage{float}

\title[Fast and mildly relativistic shock breakout]{Monte-Carlo simulations of  fast Newtonian  and mildly relativistic shock breakout from a stellar wind}
\author[Ito et al.]{{Hirotaka Ito$^{1,2}$, Amir Levinson$^{3}$ and
    Ehud Nakar$^{3}$}\\ 
  $^{1}$Astrophysical Big Bang Laboratory, RIKEN, Saitama 351-0198, Japan; hirotaka.ito@riken.jp\\
  $^{2}$Interdisciplinary Theoretical \& Mathematical Science Program (iTHEMS), RIKEN, Saitama 351-0198, Japan\\
$^{3}$School of Physics \& Astronomy, Tel Aviv University, Tel Aviv 69978,
  Israel; levinson@wise.tau.ac.il\\
  }

\begin{document}
\date{\today}
\pagerange{000--000} \pubyear{0000}
\maketitle
\label{firstpage}
\begin{abstract}
Strong explosion of a compact star surrounded by a thick stellar wind drives a fast ($>0.1c$) radiation mediated shock (RMS)
that propagates in the wind, and ultimately breaks out gradually once photons start escaping from the shock transition layer.    
In exceptionally strong or aspherical  explosions the shock velocity may even be relativistic.   The properties of the breakout signal depend on the dynamics and
structure of the shock during the breakout phase.   
Here we present, for the first time, spectra and lightcurves of the breakout emission of fast Newtonian
and mildly relativistic shocks, that were calculated using self-consistent Monte-Carlo simulations of finite 
RMS with radiative losses.  We find a strong dependence of the $\nu F_\nu$ peak on shock velocity, ranging from $\sim 1$ keV
for $v_s/c=0.1$ to $\sim 100$ keV for $v_s/c=0.5$, with a shift to lower energies as losses increase.
For all cases studied the spectrum below the peak exhibits
a nearly flat component ($F_\nu \sim \nu^0$) that extends down to the break frequency below which absorption becomes important.  
This implies much bright optical/UV emission than hitherto expected.   The computed lightcurves 
show a gradual rise over tens to hundreds of seconds for representative conditions.  The application to SN 2008D/XRT 080109 and the detectability limits are 
also discussed.  We predict a detection rate of about one per year with eROSITA.
\end{abstract}


\section{Introduction}
\label{Intro}

The collapse of a massive star creates a radiation dominated shock wave that propagates in the stellar envelope, breaks out, and ultimately emits the observed supernova light.  In the majority of core-collapse events the breakout occurs at the edge of the stellar envelope, however, in stars that eject a sufficiently intense stellar wind prior to their collapse  the RMS continues to propagate in the wind until reaching a large enough radius at which breakout ensues
\citep{campana2006,S08,waxman2007,KBW10,balberg2011,CI2011,CI2012,LN19}. 
This is likely to occur in compact progenitors, like Wolf-Rayet stars, that exhibit  broad emission lines, indicating fast winds
with high mass flux.   
In particular, there is a growing body of evidence suggesting that many SN progenitors experience episodes of prodigious
mass loss shortly (months to years) before core collapse, with rates as high as $\dot{M}_w \sim10^{-3} - 10^{-1} ~M_\odot$ yr$^{-1}$ \citep{ofek2014b,galyam2014,svirski2014a}.   While more modest winds  (with $\dot{M}_w\sim 10^{-7}-10^{-4}~M_\odot$ yr$^{-1}$) 
are commonly thought to be driven by radiative pressure,
the nature of these intense eruptions is yet unclear (see discussion in, e.g., \citealt{shiode2014}).    

 If the explosion energy is high enough and/or aspherical, the shock velocity may approach the speed of light, and
in some circumstances may even be relativistic.   An example is low luminosity GRBs that, in some scenarios \citep{nakar2015}, result from the breakout
of a mildly or even highly relativistic shock from an extended envelope surrounding the compact progenitor. 
In general, a shock propagating at a velocity $v_{s}/c > \tau_{w\star}^{-1} = 10^{-2} (R_{\star11}v_{w3}/\kappa_{0.2}\dot{M}_{-3})$, where
$c$ is the speed of light,
$\tau_{w*}$ is the total optical depth of the wind, $R_\star=10^{11}R_{\star11}$ cm is the progenitor radius, $v_w=10^3v_{w3}$ km/s  the wind velocity, $\dot{M}_w=10^{-3}\dot{M}_{w-3}$ the mass loss rate and
$\kappa=0.2\kappa_{0.2}$ gr$^{-1}$ cm$^{2}$ the Thomson opacity per unit mass, will remain radiation mediated upon transitioning into the 
stellar wind \citep{LN19}.  Fast Newtonian and mildly relativistic shocks, which are produced in powerful explosions of compact stars, likely 
surpass this criterion, hence their breakout is anticipated to occur gradually in the wind  (as opposed to a sudden breakout from the stellar edge), 
at radii much larger than the stellar radius (see discussion in Section \ref{LC}).   The properties of the breakout signal in such shocks is the focus of this paper.

Observational evidence for shock breakout from a wind are rare and controversial. 
The leading candidate is probably the X-ray flash from SN 2008D \citep{S08,modjaz2009}, for which the various properties of the emission suggest that the breakout occurred in a dense stellar wind rather than from the surface of the progenitor \citep[e.g][]{S08,balberg2011,svirski2014a}.  Another type of SNe in which the emission is associated with a shock propagating in a wind are type IIn SNe, that show a bright and blue light curve and are thought to be powered by interaction.  The rise time in these SNe  has been attributed to a shock breakout emission \citep[e.g.,][]{ofek2010,ofek2014}. The last type of SNe that were suggested to be a breakout through a stellar wind are bright and very long ultra-luminous SNe where the mass of the wind is so large (several solar masses or more), that the breakout signal constitutes practically the entire main part of the SN light. The prototype of this class is SN2006gy  \citep[e.g.,][]{CI2011}.

During the gradual breakout from the wind the radiative losses continuously increase \citep{GNL18,LN19}.   These losses can significantly alter 
the shock structure and emission.   For sufficiently slow shocks ($v_s/c \lesssim 0.05$) the radiation is in full thermodynamic equilibrium already
inside the shock transition layer, and the emitted spectrum is a black body spectrum.  In faster sub-relativistic shocks ($0.1 \lesssim v_s/c \lesssim 0.5$), 
termed  fast Newtonian shocks, full thermodynamic equilibrium occurs only far downstream and the immediate downstream temperature depends
sensitively on shock velocity \citep{KBW10,LN19,ILN20}.   In such shocks radiative losses can lead to notable effects.
Analytic models that invoke the diffusion approximation \citep{ILN19} 
suggest that in fast Newtonian shocks the shock thickness and the immediate downstream temperature 
decrease with increasing losses during the breakout phase.   However, what is the shape of the emitted spectrum and 
how it evolves with time is currently unknown.  Moreover, as the shock velocity approaches $0.5c$ pair 
creation may become important, further complicating the problem.
Computing the shape of the spectrum, which is particularly important 
for the estimation of detection limits at photon energies well below the X-ray peak (optical - UV in particular), is the main goal of this
paper.   Indeed, we show below that the flux in the optical-UV band is vastly higher (by several orders of magnitude) than that anticipated assuming 
a Wien spectrum. 

The analysis outlined in this paper exploits a modified version of our Monte-Carlo code  \citep[][hereafter ILN20]{ILS18,ILN20} that incorporates photon escape,
thereby enabling the calculations of the structure and spectrum of RMS during the breakout phase, when losses become substantial. 
The model tacitly assumes that the shock evolves in a quasi-steady manner, in the sense that it adjusts to the local conditions at
any time such that a steady-state solution provides a good approximation to its structure and emission. 
The range of  shock velocities considered here  encompasses the fast Newtonian to mildly relativistic regimes, $0.1 \leq \beta_{\rm u} \leq 0.5$,
where $\beta_{\rm u}$ is the velocity of the upstream plasma in units of the speed of light, as measured in the shock frame.
Fully relativistic RMS, that possess vastly different properties, will be considered in a follow up paper.

 This paper is organized as follows.   In Section \ref{Sec:numerical} we describe the numerical method and the setup of  our simulations.
 We present the computed structure of RMS in Section \ref{Sec:structure}. The resulting spectrum of the photons escaping from RMS is shown in Section \ref{Sec:spectrum}.
 Lightcurves computed by combining the simulations results with a realistic shock propagation model are presented in section \ref{LC}.
 Applications to SN 2008D/XRT 080109 are discussed in section \ref{sec:comparison} and detectability considerations in Section \ref{sec:Detect}.
 We conclude in Section \ref{Sec:conclusion}.
Throughout the paper, the subscript $u$ and $d$  refer to the physical quantities at the far upstream and 
far downstream  regions of the shock, respectively.

\section{Numerical Setup}
\label{Sec:numerical}
%

The details of  the numerical method  are described in \citet{ILS18} and ILN20,
where calculations of infinite shocks, i.e., shocks of sufficiently large optical depth that prevents any radiative losses,
are presented.  Here we extend the calculations to shocks of finite optical depth that allow photon escape from the upstream boundary of the shock.

In computing finite shock solutions, we employ two different methods, depending on whether a subshock forms or not.
Our strategy to compute a smooth finite shock profile (that do not sustain a subshock) is to fix the downstream velocity  $\beta_{\rm d}$ at a value 
lower than that obtained for an infinite shock with the same upstream conditions.
For given downstream conditions, our code iteratively seeks  a 
steady shock profile  that conserves the energy-momentum flux throughout the flow, as in the case of an infinite shock
\citep[for further details, see][ILN20]{ILS18}.
Since the compression ratio of the shock increases when radiative losses are present,
this lower $\beta_{\rm d}$ leads to a solution with a larger energy escape.
In any case, we first apply the above method to compute the finite shock structure.
If the code fails to achieve convergence in this way, we introduce a subshock in the flow.
In this alternative approach, we no longer fix the downstream velocity, instead the optical depth from the upstream boundary to the position of subshock, $\tau_{\rm sub}$, is fixed during the iteration.
For the given optical depth  $\tau_{\rm sub}$, our code again seeks a steady profile with a subshock.\footnote{As found in the case of infinite shocks \citep[][ILN20]{ILS18}, the error in the energy-momentum conservation condition along the flow is reduced (converges within few percent) by introducing a subshock in the system when the code fails to find a smooth solution.}
In this case, a solution with larger energy escape can be obtained by reducing the value of $\tau_{\rm sub}$.
Note that the downstream velocity $\beta_{\rm d}$ is obtained as an eigenvalue in this approach.
In principle, it is possible to iteratively seek  a solution with a subshock by fixing $\beta_{\rm d}$ and treating $\tau_{\rm sub}$ as a free variable in the iterations.
However,  since extremely high spacial resolution is required around the subshock in order to accurately resolve the flow profile there, and since 
the position of the subshock is unknown a priori, it renders this technique far more challenging,

The radiative losses are quantified by the escape parameter $f_{esc}$ 
which is the ratio of the energy flux carried by the escaping photons
to the incoming energy flux of the baryons far upstream:
\begin{eqnarray}
\label{eq:fraction}
f_{esc} = - \frac{F_{esc}}{F_{b}}. 
\end{eqnarray}
Here $F_{esc}<0$ denotes the net energy flux of the photons at the upstream boundary of the simulation box\footnote{$F_{esc}$ is a negative quantity since we define positive energy  flux in the direction along the flow.}  and
$F_{b} = \Gamma_{\rm u}(\Gamma_{\rm u} - 1)n_{\rm u}m_p \beta_{\rm u} c^3$ is the energy flux of the baryons, where 
$m_p$ 
is the proton rest mass and $\Gamma_{\rm u}=(1-\beta_{\rm u}^2)^{-1/2}$  is the Lorentz factor of the upstream flow.
Note that in our approach the value of $f_{esc}$ is not an input parameter but rather an output of the
calculations.
A set of RMS solutions with different values of $f_{esc} $ is obtained below by performing many simulations with 
different input values ($\beta_{\rm d}$ for smooth solutions and $\tau_{\rm sub}$ when a subshock is present).

In addition to $\beta_{\rm d}$ or $\tau_{\rm sub}$, the input parameters of the simulations are
the velocity of the upstream flow, $\beta_{\rm u}$, the proper baryon density at the far upstream region  $n_{\rm u}$, and the composition which we take to be purely hydrogen. 
As stated in the introduction, we are interested in exploring the regimes of fast Newtonian and mildly relativistic shocks.
To that end we consider 3 models with different values of the upstream velocity, $\beta_{\rm u} = 0.1, 0.25$ and $0.5$.
As for the baryon density, we invoke a fixed value of $n_{\rm u} = 10^{15}~{\rm cm}^{-3}$ for the fiducial models
which is identical to that adopted in the calculations of infinite shocks presented in ILN20.  
To explore the dependence on the density,  we also compute a subset of RMS solutions 
with $n_{\rm u} = 10^{12}~{\rm cm}^{-3}$  for each $\beta_{\rm u}$.

\section{The structure of RMS with escape}
\label{Sec:structure}

As shown in our previous study (ILN20), the properties of infinite RMS are  vastly different in the sub-relativistic and  the relativistic regimes.  
This holds true also for finite shocks.  Below, we discuss the properties of finite sub-relativistic RMS  ($\beta_{\rm u} = 0.1$ and $0.25$) and 
mildly relativistic RMS  ($\beta_{\rm u} = 0.5$) separately.

\subsection{Sub-relativistic RMS}
\label{Sec:subrela}

\begin{figure*}
\begin{center}
\includegraphics[width=17.5cm,keepaspectratio]{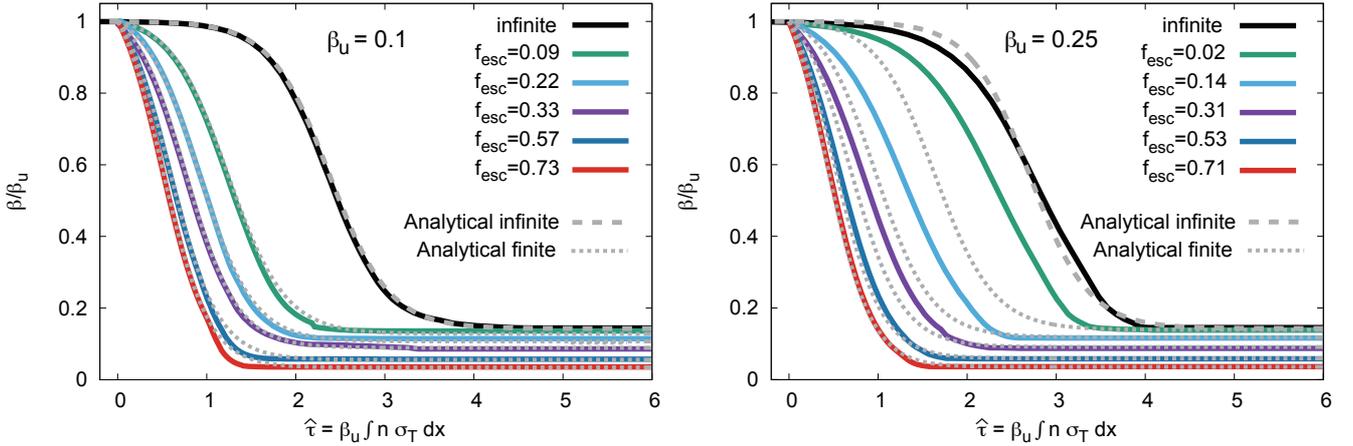} 
\end{center}
\caption{Velocity profiles of RMS with upstream velocities  $\beta_{\rm u} = 0.1$ ({\it left}) and  $\beta_{\rm u} = 0.25$ ({\it right}), plotted
   as functions of the normalized optical depth $\hat{\tau} = \beta_{\rm u} \int n \sigma_T dx$. The solid black
   line in each panel displays the simulation result for the infinite shock, while the green, cyan, magenta, blue and red lines depict the results for finite shocks with downstream velocities  $\beta_{\rm d}/ \beta_{\rm d, inf} = 0.95$, $0.8$, $0.6$, $0.4$ and $0.25$, respectively, where $\beta_{\rm d, inf}$ denotes the downstream velocity of the infinite shock ($f_{esc} = 0$).  The resulting escape fraction obtained in each simulations
  is indicated in the figure legends.  The grey dashed line marks the analytical infinite shock solution and the
  grey dotted lines are the corresponding analytical finite shock solutions
  obtained for the values of $f_{esc}$ and $\alpha$ found in the simulations.
  The values of $\alpha$ used for the fits are $14.2$ and $5.6$ for $\beta_{\rm u} = 0.1$ and $0.25$, respectively, independent of $f_{esc}$.}
\label{vsub}
\end{figure*}

In Fig. \ref{vsub}, we plot the velocity profiles
of infinite and finite fast Newtonian RMS ($\beta_{\rm u} = 0.1$ and $0.25$)
as a function of the normalized optical depth, defined as  $\hat{\tau} = \beta_{\rm u} \int n \sigma_T dx$,  
for a wide range of escape fractions, up to $f_{\rm esc} \sim 0.7$.
For this range of $f_{\rm esc}$, all solutions were found to have a smooth profile without a subshock.
For an infinite shock, the width of the shock transition layer, $l_{sh}$
is determined by the diffusion length of the photons ($\sigma_T n_{\rm u} l_{sh} \approx 1/\beta_{\rm u}$).
Photon leakage is anticipated when the optical depth of the shock becomes smaller than this value.
Consequently, as $f_{esc}$ increases the shock width is expected to become narrower.  This trend is clearly seen in Fig \ref{vsub}.

In Fig \ref{vsub} we also plot  the analytic RMS solutions (dotted lines) derived using the model outlined in \citet{ILN19}.
These analytic solutions are characterized by a dimensionless free parameter that fixes the radiation pressure 
at the upstream boundary.  It is given explicitly as:
\begin{eqnarray}
\label{eq:alpha}
\alpha =  \frac{f_{esc}}{2 p_{esc}},
\end{eqnarray}
where $p_{esc} = P_{esc} / (\Gamma_{\rm u}^2 n_{\rm u} m_p \beta_{\rm u}^2 c^2)$ is the momentum flux of the 
photons normalized by the baryon momentum flux at the upstream.
For each of the analytic solutions depicted in Fig \ref{vsub}, we  adopted an $\alpha$ value 
that was self-consistently determined from the simulation with same escape fraction.
Interestingly, we find that for a given choice of  $\beta_{\rm u}$ the value of alpha thereby obtained
is independent of the escape fraction $f_{esc}$.
For $\beta_{\rm u} = 0.1$ ($0.25$), the finite shock simulations yield
$\alpha = 14.2$ ($5.6$), with less than $1\%$ deviation, for all the cases explored in the current study.

As already shown in our previous paper, there is excellent agreement between the numerical and analytical solutions 
of an infinite shock with  $\beta_{\rm u} = 0.1$.
As for an infinite shock with $\beta_{\rm u} = 0.25$, the analytic solution was also found to be in good agreement with the simulations, albeit 
with notable (though small) deviations owing to the larger inaccuracy of the diffusion approximation in this case.
The finite shock solutions are also in good agreement with the simulations, with nearly perfect match for $\beta_{\rm u} = 0.1$ and 
larger deviations for $\beta_{\rm u} = 0.25$.
This confirms that the diffusion approximation is reasonable for fast Newtonian RMS even in the presence of large radiative losses.

\begin{figure*}
\begin{center}
\includegraphics[width=17.5cm,keepaspectratio]{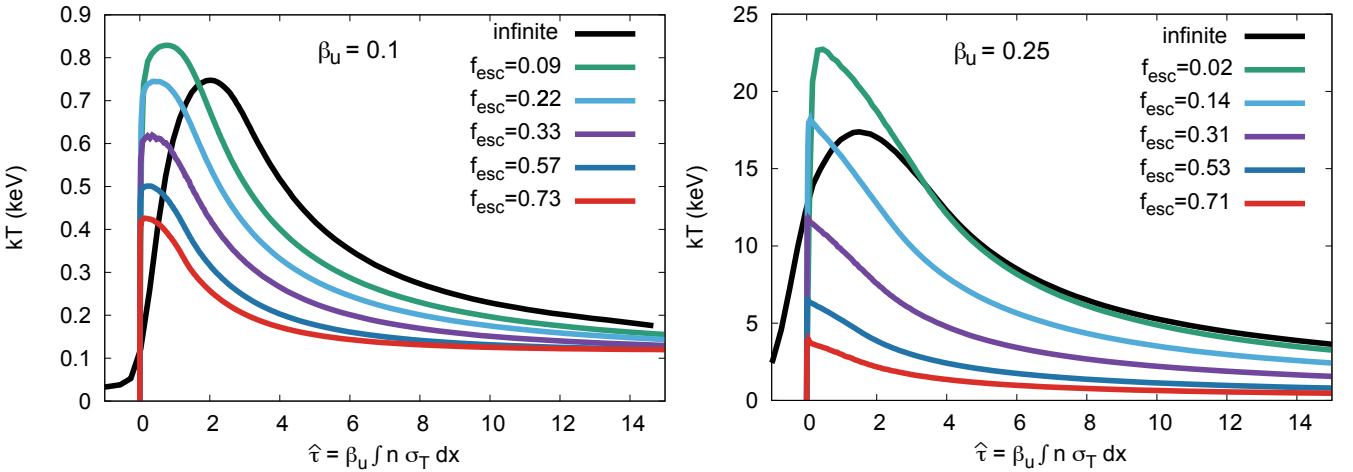} 
\end{center}
\caption{Temperature profiles obtained for the solutions in Fig. \ref{vsub}. Note that the extent of the horizontal axis is larger than in Fig. \ref{vsub}.
}
\label{Tsub}
\end{figure*}

Fig. \ref{Tsub} exhibits the corresponding temperature profiles.
The sensitive dependence of the temperature on the upstream velocity seen in the figure
is consistent with previous findings for infinite shocks \citep[][ILN20]{W76,KBW10}
\footnote{From the current simulations we find that the temperature roughly scales as $T\propto \beta_{\rm}^{3.4}$ in the range $0.1 \leq \beta_{\rm u} \leq 0.25$.
  This is slightly steeper than the dependence $T \propto \beta_{\rm u}^3$ found in ILN20 
  in the range $0.1 \leq \beta_{\rm u} \leq 0.5$, since the regulation of temperature by the vigorous pair production is already important at $\beta_{\rm u} = 0.5$.}.
The decline of the temperature with increasing losses (larger values of $f_{esc}$) is consistent with the trend found in \cite{ILN19}.
The reason for this behaviour is that larger losses give rise to a higher compression ratio (i.e., a smaller downstream velocity) and, consequently, 
a larger diffusion length behind the shock which, in turn, enhances photon production in the immediate downstream.
Our simulations confirm that the decline in temperature during a gradual breakout  is a robust feature.

\subsection{Mildly relativistic RMS}
\label{Sec:subrela}
\begin{figure}
\begin{center}
\includegraphics[width=8.2cm,keepaspectratio]{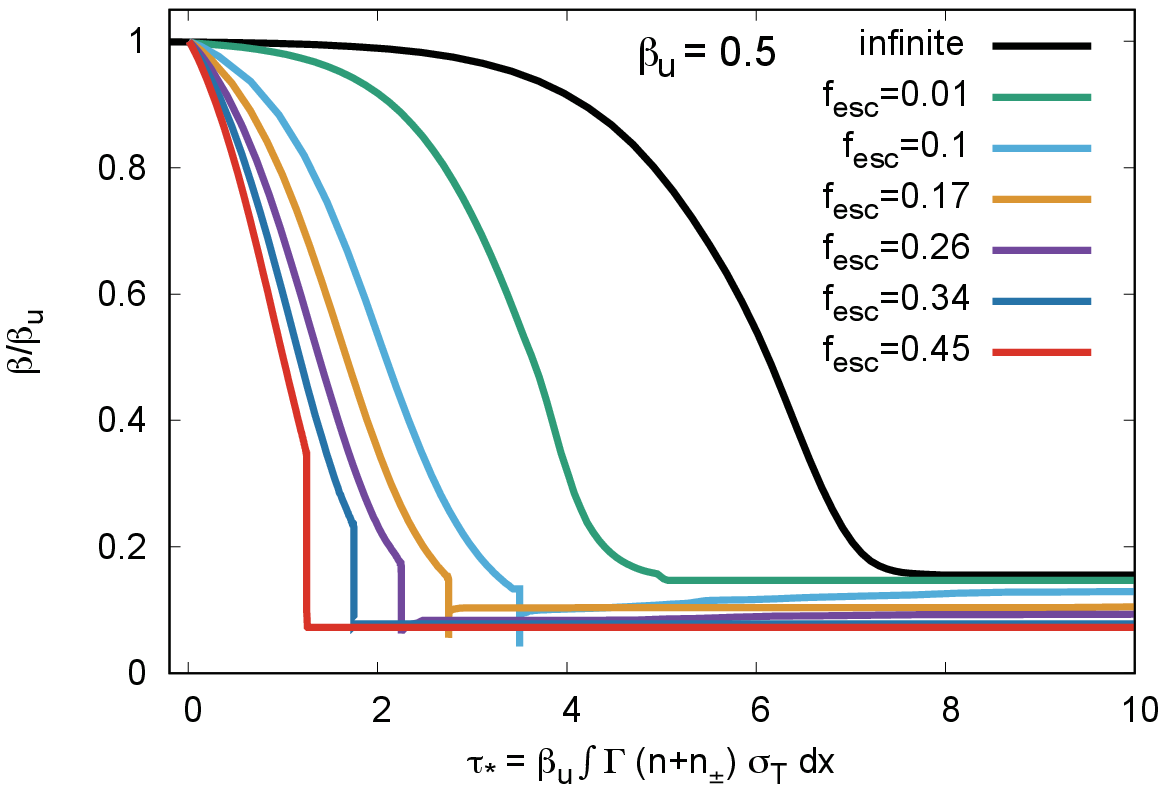} 
\end{center}
\caption{Velocity profiles of RMS with an upstream velocity  $\beta_{\rm u} = 0.5$, plotted
  as functions of the normalized pair-loaded optical depth $\tau_* = \beta_{\rm u} \int \Gamma (n+n_{\pm}) \sigma_T dx$.
  The profiles depicted by the solid black (infinite shock) and green lines are found to be smooth (contain no subshock). 
  The  cyan, brown, magenta, blue and red lines show the simulation results for finite shocks that contain a subshock.
  The resulting escape fractions are given in the figure legends.
 }
\label{vB5}
\end{figure}

\begin{figure}
\begin{center}
\includegraphics[width=8.2cm,keepaspectratio]{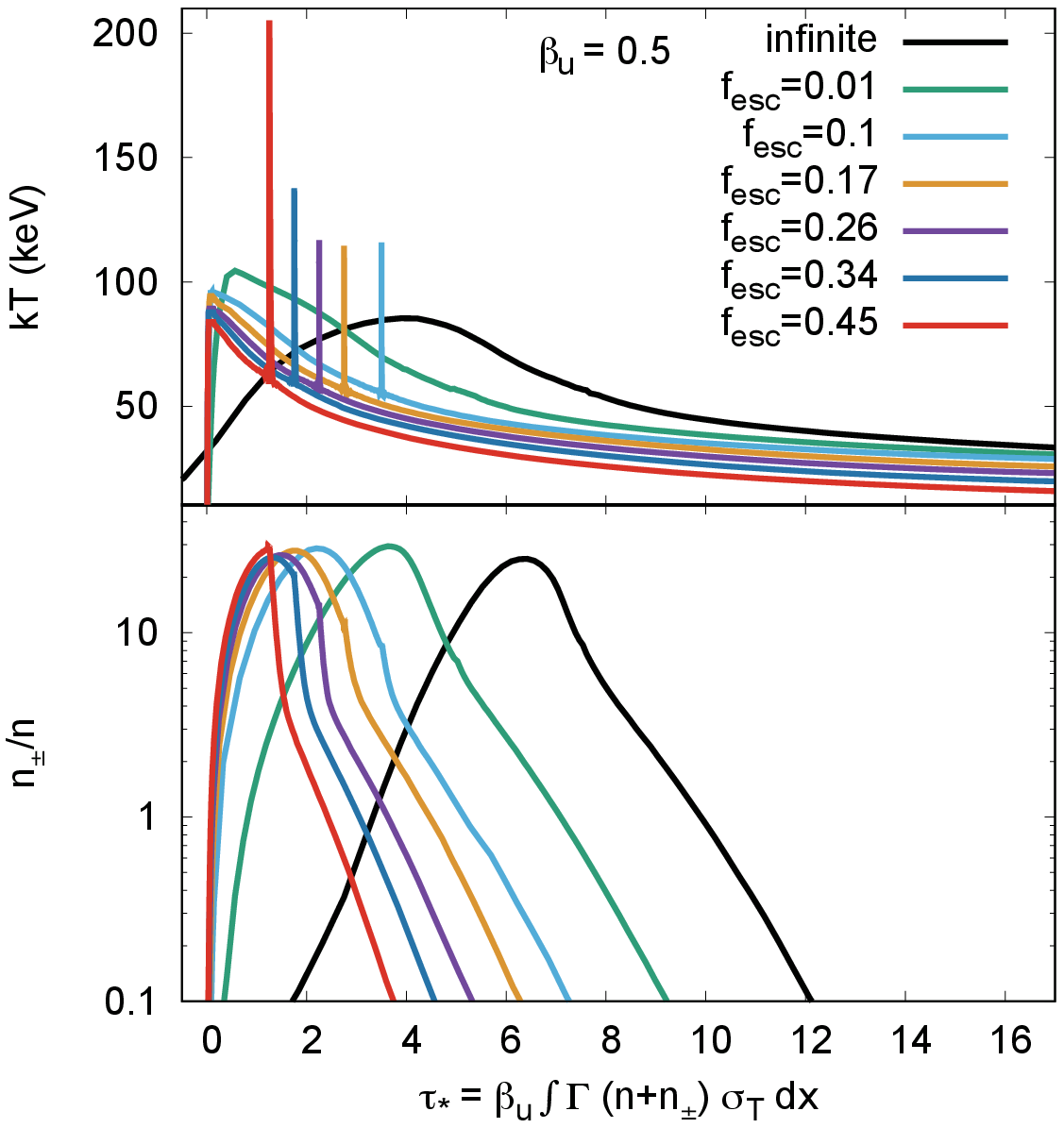} 
\end{center}
\caption{Profiles of temperature ({\it top}) and pair-to-baryon ratio ({\it bottom}) obtained for the solutions shown in  Fig. \ref{vB5}.
Note the larger extent of the horizontal axis.
}
\label{TnB5}
\end{figure}

The velocity profiles of RMS with $\beta_{\rm u} =0.5$ and different escape fractions are displayed in Fig. \ref{vB5}.
In this regime the diffusion approximation adopted in \cite{ILN19}  is totally inapplicable, hence analytic solutions
cannot be obtained.  Moreover, the shock opacity is dominated by newly created pairs (Fig. \ref{TnB5}.) and, therefore, the 
solutions are given as functions of the normalized pair loaded optical depth,  $\tau_* = \beta_{\rm u} \int \Gamma (n+n_{\pm}) \sigma_T dx$.

As in the previous cases, the shock transition layer becomes narrower as the escape fraction increases.  On the other hand, unlike 
the previous cases, solutions with sufficient losses (more than a few percents) exhibit a subshock.  As seen in the
figure, the strength of the subshock increases with increasing $f_{esc}$, becoming quite large as $f_{esc}$ approaches $0.5$.
%

The profiles of the temperature and pair-loading parameter (pair-to-baryon ratio) are shown in Fig. \ref{TnB5}.
As seen, unlike in fast Newtonian shocks, the immediate downstream temperature in this case is practically 
independent of  $f_{esc}$.  This is a consequence of temperature regulation by exponential pair creation
(for a detailed explanation of this effect see \citealt{LN19}, and references therein).
The spikes seen in the temperature curves correspond to overheated plasma immediately behind  the subshock.   
Since the subshock is collisionless, heating of the plasma occurs 
on kinetic scales which are vanishingly small.  The width of the spike is thus determined by the cooling 
length of the overheated plasma, which is much smaller than the photon mean free path, as explained in detail in \cite{ILS18}.

\subsection{Dependence of shock structure on the upstream density}
\label{Ndep1}
\begin{figure*}
\begin{center}
\includegraphics[width=17.5cm,keepaspectratio]{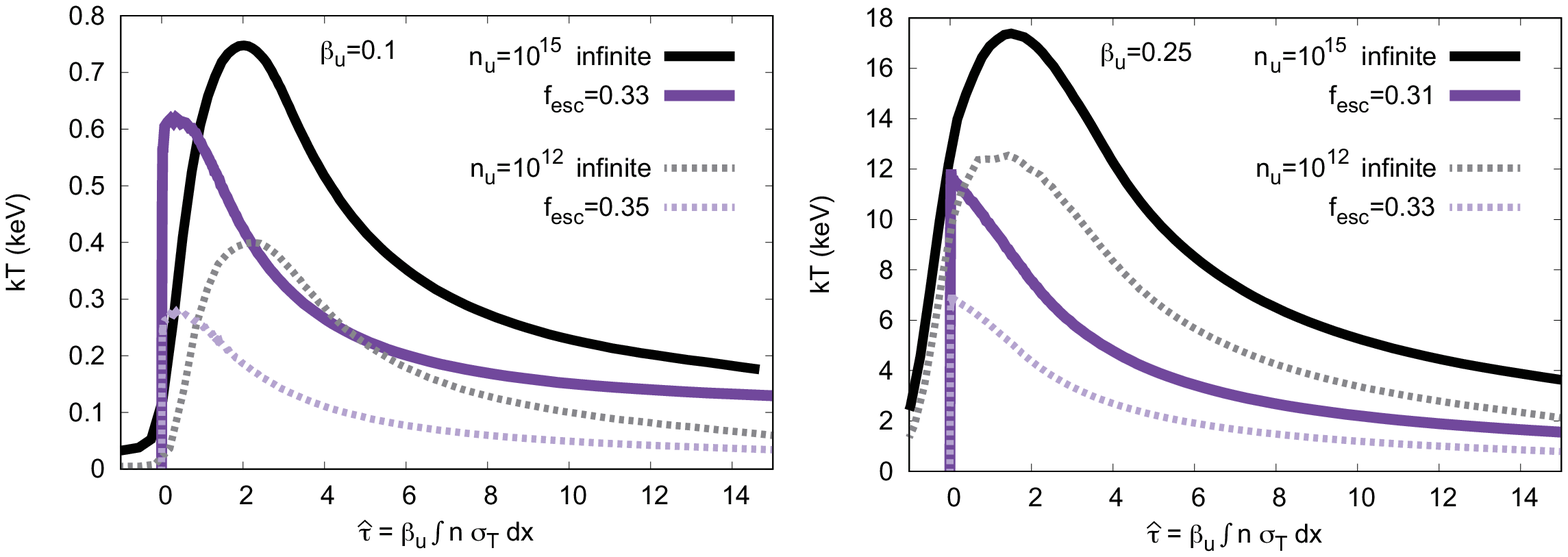} 
\end{center}
\caption{The temperature profile 
  for the simulations with upstream density of $n_{\rm u} = 10^{15}~{\rm cm}^{-3}$ ({\it solid lines}) and $10^{12}~{\rm cm}^{-3}$ ({\it dotted lines}).
    The left and right panels display the results for $\beta_{\rm u} = 0.1$ and $0.25$, respectively.
  The black and magenta lines show the cases of the infinite and finite shocks for $\beta_{\rm d} = 0.6 \beta_{\rm d, inf}$, respectively.
  Note that the slight difference found in the escape fraction of finite shocks 
  for different $n_{\rm u}$ is  due to the  numerical errors.
}
\label{B1ndep}
\end{figure*}

\begin{figure}
\begin{center}
\includegraphics[width=8.2cm,keepaspectratio]{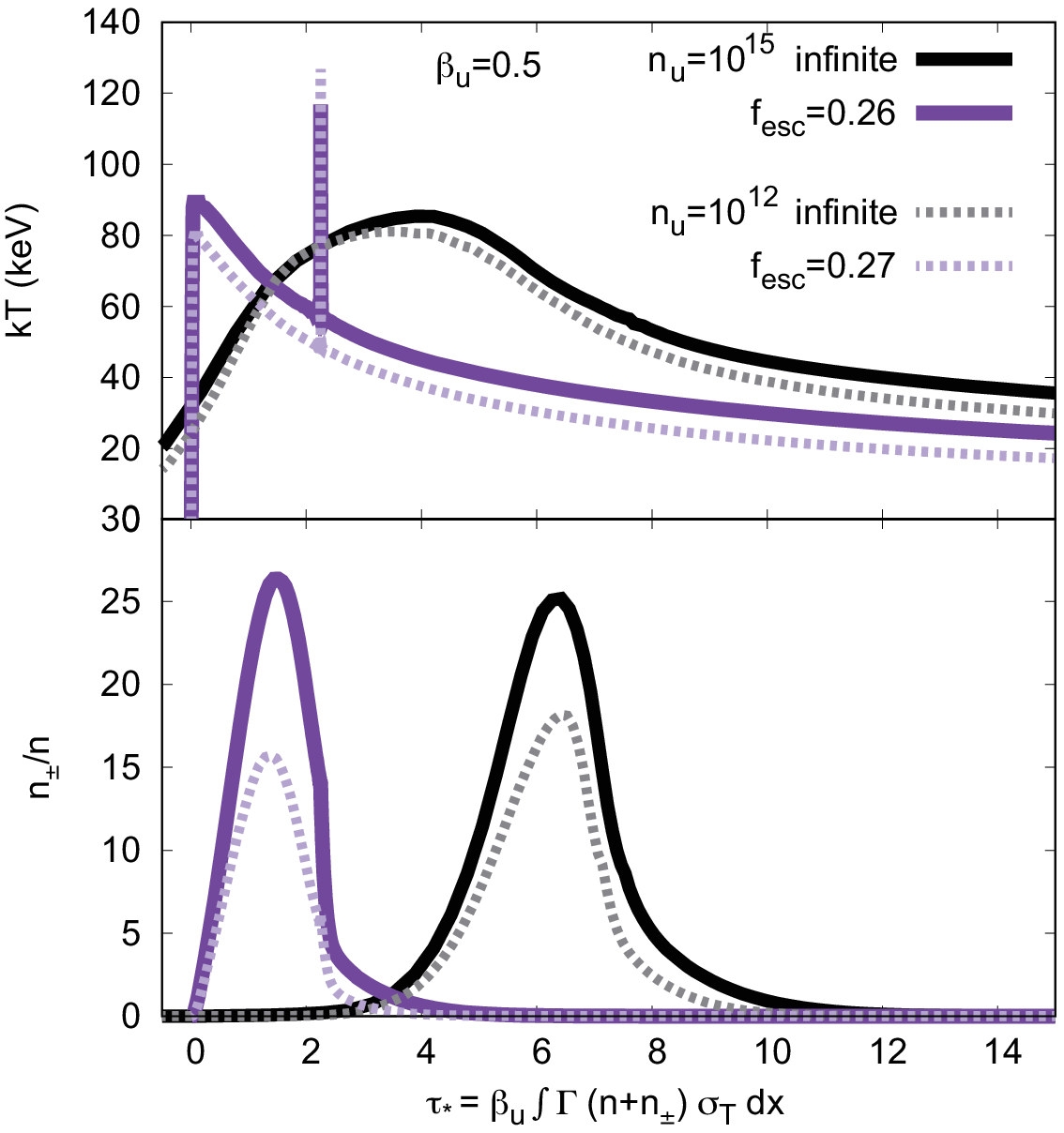} 
\end{center}
\caption{The temperature ({\it top}) and pair-to-baryon ratio ({\it bottom}) profiles for the simulations with upstream density of $n_{\rm u} = 10^{15}~{\rm cm}^{-3}$ ({\it solid lines}) and $10^{12}~{\rm cm}^{-3}$ ({\it dotted lines}).
  The black and magenta lines show the cases of the infinite and finite shocks for $\tau_{\rm sub} = 2.25$, 
  respectively.
  Note that the slight difference found in the escape fraction of finite shocks 
  for different $n_{\rm u}$ is  due to the  numerical errors.}
\label{B5ndep}
\end{figure}

In all the above solutions the upstream baryon density was taken to be $n_{\rm u} = 10^{15}~{\rm cm}^{-3}$.
This particular value was chosen in order to compare the finite shock solutions with the  simulations of infinite shocks 
performed by ILN20.  However, in most cases of shock breakout from a stellar wind the typical density is expected to
lie in the range $n_{\rm u} \sim 10^{12}-10^{13}~{\rm cm}^{-3}$.
While the velocity profile as a function of optical depth is independent of $n_{\rm u}$, the temperature (as well as the pair density for $\beta_{\rm u} = 0.5$)
and the photon spectrum do have certain dependences on the number density.
To elucidate the dependence of RMS properties on the upstream baryon density we performed additional simulations 
of finite and infinite shocks with $n_{\rm u}=10^{12}~{\rm cm}^{-3}$.
In Fig. \ref{B1ndep} we compare temperature profiles of sub-relativistic shocks ($\beta_{\rm u}=0.1$ and $0.25$) obtained
for $n_{\rm u}=10^{15}~{\rm cm}^{-3}$ and $n_{\rm u}=10^{12}~{\rm cm}^{-3}$.
As seen, a lower density gives rise to a lower temperature, although the dependence is rather weak (a factor of 2 change over three decades in density). 
This, nonetheless, has important  impact on the spectral luminosity at frequencies below the peak, as will be discussed in section \ref{LC}.
The results exhibited in Fig. \ref{B1ndep}  are in very good agreement with analytic estimates \citep{ILN19,LN19}.
A similar comparison for the $\beta_{\rm u} = 0.5$ shock is exhibited in 
Fig. \ref{B5ndep}, and it is seen that in this case
the temperature is practically independent of density, whereas the dependence of the pair content is very weak. 
This is a consequence of the pair creation thermostat discussed above.

\section{The spectrum of escaping radiation}
\label{Sec:spectrum}

\begin{figure}
\begin{center}
\includegraphics[width=8.2cm,keepaspectratio]{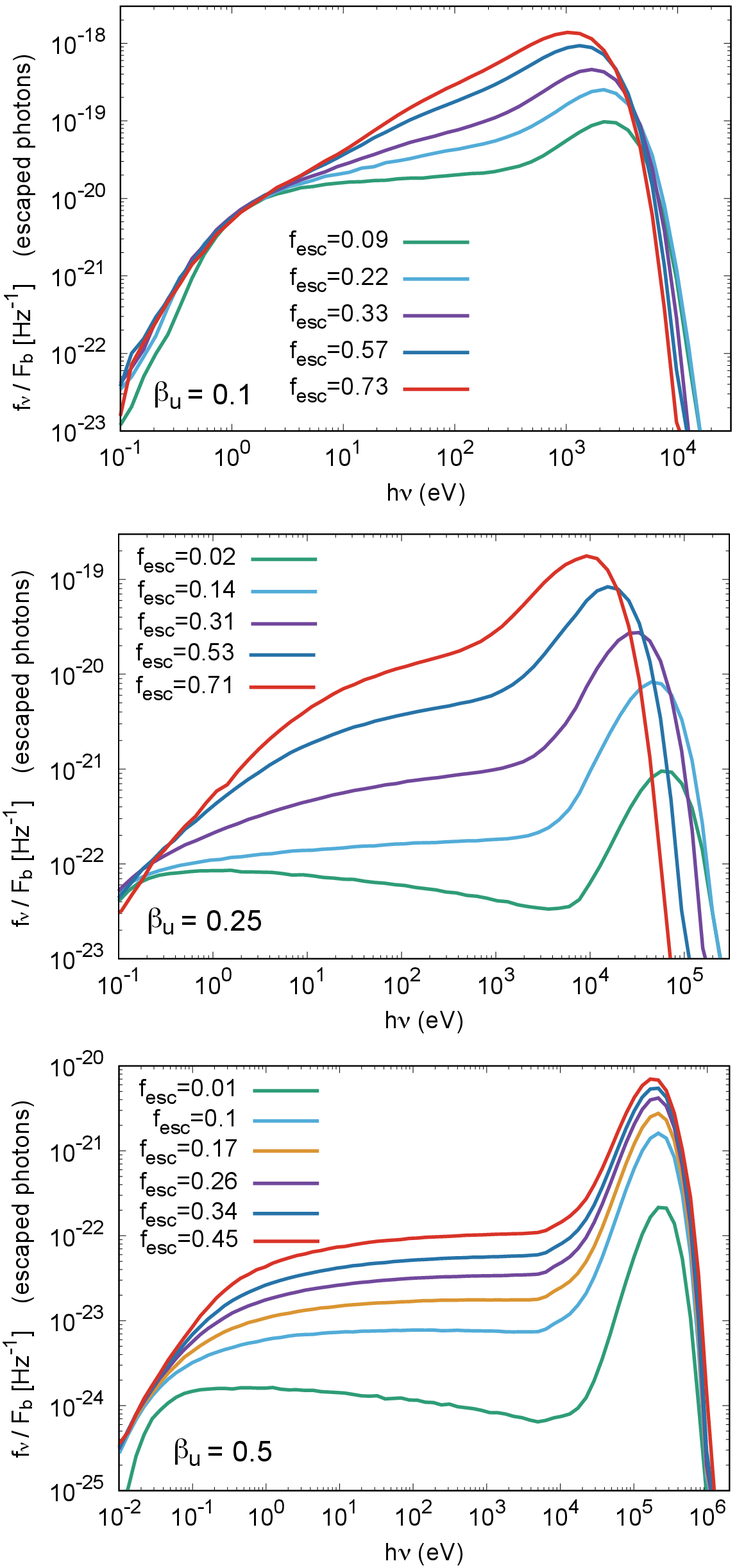}
\end{center}
\caption{Shock-frame, flux density of escaping photons, $f_{\nu} = - \frac{dF_{esc}}{d\nu}$, normalized by the total kinetic energy flux of baryons at the upstream boundary, $F_b = \Gamma_{\rm u} (\Gamma_{\rm u} - 1) n_{\rm u} m_p c^3 \beta_{\rm u}$. 
  The top, middle and bottom panels display the results for shock velocities $\beta_{\rm u} = 0.1$, $0.25$ and $0.5$, respectively.
  The different lines in each panel correspond to different values of the escape fraction, as indicated.
}
\label{SPDS}
\end{figure}

In Fig. \ref{SPDS},
we show the spectral energy distribution of  photons escaping through the upstream boundary of the simulation box, $f_\nu = - \frac{dF_{esc}}{d\nu}$,
for each model.    Each line, computed for a particular value of $f_{esc}$, as indicated in the figure,  
represents the instantaneous spectrum emitted during a gradual shock breakout from a wind-like medium  at the radius at which 
the optical depth to infinity ahead of the shock roughly equals the local shock width (smaller $f_{esc}$ values correspond to earlier emission).
However, the overall normalization of the spectra does not take into account the full evolution of the shock and the structure of the ejecta.
Lightcurves computed by combining the simulations results with a realistic shock propagation model are presented in section \ref{LC}. 
It is emphasized that our results provide, for the first time,
spectra of shock breakout emission in a wind
from self-consistent simulations.

In all cases exhibited in Fig. \ref{SPDS}, the spectral peak energy reflects the immediate downstream temperature, that is, $E_p \sim 3 k T_d$.
The spectral softening (i.e., the shift of $E_p$ to lower energies) during the rise of the luminosity seen in the Newtonian shocks is a consequence of the
decline of the downstream temperature (see Fig. \ref{Tsub}).   
For a shock with $\beta_{\rm u} = 0.1$ ($0.25$), the spectral peak evolves in the soft, $E_p \sim~1{\rm keV}$ (hard, $E_p\sim 10~{\rm keV}$),
X-ray band.
As discussed in \citet{ILN19}, the superposition of emission during 
the hard-to-soft evolution may account for the time integrated, non-thermal spectrum 
observed in the shock breakout candidate XRT080109 \citep{S08}.
In contrast to the fast Newtonian shocks, the mildly relativistic ($\beta_{\rm u}=0.5$) shock shows no softening,
with $E_p$ maintained around $\sim 200~{\rm keV}$ during the luminosity rise.  This is again a consequence of the 
pair thermostat mentioned in the previous section.

A notable feature common to all spectra is the sudden change is slope below the peak.  While the portion of the spectrum around the peak (the bump) 
has a Wien shape ($I_\nu \propto \nu^3 {\rm exp}(-h\nu/kT)$) \footnote{This bump is more prominent in faster shocks, for which the departure from
thermodynamic equilibrium is larger.  For the cases studied here we find that the Wien spectrum provides 
a good fit for the entire bump only for $\beta_{\rm u}=0.25$ and $0.5$. 
For $\beta_{\rm u} = 0.1$  it  can only  fit the spectral portion above the peak ($h\nu \gtrsim E_p$).}, 
the soft tail below the peak has a spectral slope close to that of 
free-free emission,  $I_\nu \propto \nu^0$, extending down to the break frequency below which the free-free absorption is fast 
enough to establish a full thermodynamic equilibrium (and the spectrum hardens to a black body slope). 
%
%
As discussed in ILN20, 
the existence of a substantial soft tail implies that the breakout signal well below the spectral peak should be much brighter (by orders of magnitude) than
that  naively expected by invoking a Wien spectrum in the entire spectral range.
This has important implications for detection limits in optical/UV band and the interpretation of shock breakout signals (see section \ref{LC}
for detailed calculations).
%

Regarding the spectral portion above the peak, we find no notable deviations from a Wien spectrum.
Hence,  an exponential cut-off at high energies is likely to be a robust feature of (planar) fast Newtonian and, perhaps, mildly relativistic shocks at breakout.
Note, however, that following the breakout episode the shock transforms into a collisionless shock that keeps propagating in the 
optically thick medium; during this phase a power-law spectrum is expected, as discussed in \cite{svirski2014a}.
The results of our simulations indicate that, contrary to previous claims \citep{WWM07, SS10},   bulk Comptonization is unlikely to be 
the origin of the high energy, non-thermal tail observed in XRT080109. 
It is worth noting that for the $\beta_{\rm u} = 0.5$ shock, slight hardening of the spectrum may occur when the losses exceed 
the values explored here ($f_{esc} = 0.45$).  
Such deviations are indeed indicated by preliminary calculations with larger escape fractions.
However, we find that when  $f_{esc} > 0.45$, the subshock becomes exceedingly strong  and intermittent, and the simulation
does not converge to a steady-state solution.\footnote{To be concrete, for any number of iterations, the simulation cannot find a steady profile which
satisfies energy-momentum conservation to within an error of a few $\%$.}
This might suggests that  the transition to the  collisionless regime becomes fully dynamic, likely involving turbulence and other stochastic effects.
In this regard it is worth pointing out that the strong subshock may give rise to efficient particle acceleration.  Once the energy dissipated 
in the subshock amounts to a considerable fraction of the total shock energy, Compton scattering and synchrotron emission by the accelerated
pairs may significantly modify the high-energy portion of the spectrum, conceivably  giving rise to a nonthermal gamma-ray flash.
We defer the exploration of such effects to a future work.

\subsection{Dependence of spectrum on the upstream density}
\label{Sec:SPndep}

\begin{figure}
\begin{center}
\includegraphics[width=8.2cm,keepaspectratio]{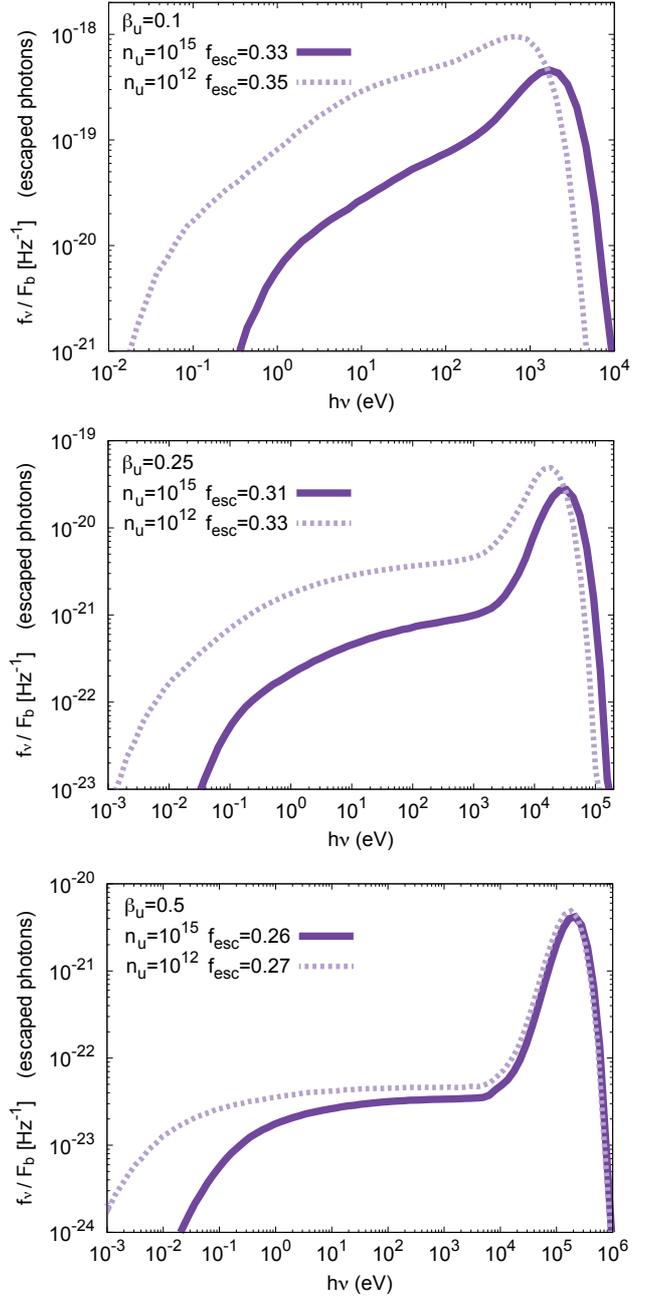} 
\end{center}
\caption{Same as Fig. \ref{SPDS}, but for comparison between $n_{\rm u}=10^{15}~{\rm cm}^{-3}$ ({\it solid lines}) and  $n_{\rm u}=10^{12}~{\rm cm}^{-3}$ ({\it dotted lines}) at a nearly identical escape fraction.
}
\label{SPndep}
\end{figure}

A comparison of spectra obtained for shocks with upstream densities  $n_{\rm u}=10^{15}~{\rm cm}^{-3}$  and $n_{\rm u}=10^{12}~{\rm cm}^{-3}$ 
is given in Fig. \ref{SPndep}.  
As seen, the main effect is a shift of the spectrum to lower energies as the upstream density decreases, with a little change in the overall spectral shape. 
This behaviour stems from the dependence of the downstream temperature on density (see Figs. \ref{B1ndep} and \ref{B5ndep}).
The shift is smaller the larger the shock velocity is, and is practically absent in the $\beta_{\rm u}=0.5$ case.  

An important consequence of this dependence is that the relative brightness of emission
below the peak increases substantially with decreasing
density (in other words, the ratio between the bolometric luminosity and the luminosity emitted in some band below the peak decreases with decreasing density).  
For example, for $\beta_{\rm u}=0.1$ ($0.25$) the ratio between the luminosity at the peak and the optical luminosity (at $\sim 1$ eV ) 
decreases by a factor of about  10 (5) as the density 
decreases from  $n_{\rm u}=10^{15}~{\rm cm}^{-3}$  to $n_{\rm u}=10^{12}~{\rm cm}^{-3}$.
Hence, shock breakout in a lower density environment is preferential for the detection of the optical/UV source.

Another effect caused by the change in density is found in the break frequency below which free-free absorption becomes important.
Since the photon density is much lower in the lower density simulation, the break occurs at a lower frequency.
Note that the large contrast in the photon number density is not apparent from the figure, since the displayed spectrum is normalized by the baryon energy flux $F_b$.

%

%

\section{LIGHT CURVE OF SPHERICAL SHOCK BREAKOUT FROM A STELLAR WIND}
\label{LC}
In this section we present approximate calculations of shock breakout lightcurves at different bands 
by combining the results of the previous section with a model for blast wave propagation in a wind.

The dynamics of the shock, and in particular the energy deposition profile, depend on the properties of the ejecta,
that should be given as input for the calculations of the shock evolution in the wind.
A common choice is the self-similar solution of  \cite{S60} that provides a good approximation for the structure of
the shocked layer near the edge of the envelope of the progenitor following the passage of shock.
%
%
The energy profile within the ejecta, obtained from the \cite{S60} solution,  
cab be expressed in terms of the ejecta velocity, $v$, as:
\begin{eqnarray}
  E(v) = E_0(v/v_0)^{-\lambda}
  = \frac{4\pi c v_0}{\kappa} R_*^2 (v/v_0)^{\lambda} ,
 \label{Epro} 
\end{eqnarray}
where  $v_0$ and $E_0 = 4 \pi c v_0 R_*^2 / \kappa$ are, respectively, the velocity and energy of the front shell of optical thickness
$c/v_0$ \citep{NS10}, and $\kappa$ denotes the opacity of the stellar envelope.
The index $\lambda$ depends on the power-law index $n_*$ of the envelope density profile near the edge as:  $\lambda = (1+0.62n_*)/0.19n_*$.
For typical envelopes $n_*=1-3$, wherein $n_*\approx 1.5$  for convective envelopes and  $n_*\approx 3$ for radiative envelopes.
Henceforth, we choose $n_*=3$  and $\kappa = 0.2~{\rm cm}^2~{\rm g}^{-1}$, which is suitable for Wolf-Rayet stars.
With this choice \citep{NS10} $\lambda=5$,
\begin{eqnarray}
  E_0 \approx 1.6\times 10^{44}~{\rm erg} \left(\frac{E_{exp}}{10^{51}~{\rm erg}}\right)^{0.58} \left(\frac{M_*}{5M_{\odot}}\right)^{-0.41} \left(\frac{R_{*}}{10^{11}~{\rm cm}}\right)^{1.66},
  \label{eq:E0}
\end{eqnarray}
and
\begin{eqnarray}
 v_0 \approx 0.3c \left(\frac{E_{exp}}{10^{51}~{\rm erg}}\right)^{0.58} \left(\frac{M_*}{5M_{\odot}}\right)^{-0.41} \left(\frac{R_{*}}{10^{11}~{\rm cm}}\right)^{-0.33},
 \label{eq:v0}
\end{eqnarray}
where $E_{exp}$ and $M_*$ denote the explosion energy and the mass of the ejecta, respectively.

In cases where the progenitor  is surrounded by an optically thick wind ($\tau_{w} > c/v_0$), 
the shock driven into the wind by the expanding ejecta remains radiation mediated.
The subsequent shock dynamics is dictated by the density profile of the wind.  We shall 
invoke a spherical wind with a density profile  $\rho_{w} \propto r^{-2}$.  The total mass swept up
by the shock as it reaches a radius $r_s$ is $m_s = \int^{r_s}_{R_*} 4\pi r^2 \rho_w dr \approx (4\pi \tau_w R_* / \kappa_w) r_s $,
where $\kappa_w$ is the opacity of the wind, henceforth assume to be equal to the envelope opacity, $\kappa_w = \kappa$, and
the  swept up energy is $E_s=m_sv_s^2$, where $v_s$ is the shock velocity at $r_s$.  Equating
$E_s$ with the energy injected into the shock by the ejecta, $E(v_s)$, yields $v_s(r_s) = v_0 (c R_* / v_0 \tau_w r_s)^{1/(\lambda + 2)}$.  We find it convenient to express the result in terms of the local optical depth, $\tau_s=\tau_w(R_*/r_s)$, rather than $r_s$. Using $\lambda=5$ and Eqs. (\ref{eq:E0}), (\ref{eq:v0})
we obtain
\begin{eqnarray}
  E_s  \approx &  1.7 \times 10^{45}~{\rm erg} & \left(\frac{E_{exp}}{10^{51}~{\rm erg}}\right) \left(\frac{M_*}{5M_{\odot}}\right)^{-0.72} \left(\frac{R_{*}}{10^{11}~{\rm cm}}\right)^{1.4} \nonumber \\ 
  && \times~ \left(\frac{\tau_{w}}{30}\right)^{1.4} \left(\frac{\tau_s}{10}\right)^{-0.72} \label{eq:SW1},
\end{eqnarray}
\begin{eqnarray}
  v_s  \approx   & 0.18 c &  \left(\frac{E_{exp}}{10^{51}~{\rm erg}}\right)^{0.5} \left(\frac{M_*}{5M_{\odot}}\right)^{-0.36} \left(\frac{R_{*}}{10^{11}~{\rm cm}}\right)^{-0.29} \nonumber \\ 
 &&   \times~ \left(\frac{\tau_{w}}{30}\right)^{-0.29} \left(\frac{\tau_s}{10}\right)^{0.14} .
\label{eq:SW2}
\end{eqnarray}
Note the very weak dependence of $v_s$ on $\tau_s$.  
The  dynamical time  can be expressed as
\begin{eqnarray}
  t = r_s / v_s \approx& 55~{\rm s}& \left(\frac{E_{exp}}{10^{51}~{\rm erg}}\right)^{-0.5} \left(\frac{M_*}{5M_{\odot}}\right)^{0.36} \left(\frac{R_{*}}{10^{11}~{\rm cm}}\right)^{1.29} \nonumber \\
  && \times~\left(\frac{\tau_{w}}{30}\right)^{1.29} \left(\frac{\tau_s}{10}\right)^{-1.14}.
  \label{eq:t}
\end{eqnarray}

A rough estimate of the breakout density, $\rho_b=\rho_w(r_b)$, where $r_b$ denotes the shock radius at breakout, can be obtained 
as follows: first we express the wind density in terms of the optical depth as $\rho_w(\tau_s)= \tau_s^2/(\kappa \tau_w R_*)$.
We then substitute the optical depth at the breakout radius, $\tau_s(r_b)=c/v_b (1+f_{\pm})^{-1}$,  into the latter expression,
where $v_b=v_s(r_b)$ and the factor $f_{\pm}$ denotes the pair-to-baryon ratio at the shock
which is only relavant for $\beta_{\rm u} = 0.5$.
This yields
\begin{eqnarray}
\rho_{b}\simeq 1.5\times10^{-10}  \left(\frac{v_b}{0.1c}\right)^{-2}  \left(\frac{\tau_{w}}{30}\right)^{-1} \left(\frac{R_{*}}{10^{11}~{\rm cm}}\right)^{-1}\left(1+f_{\pm} \right)^{-2}\quad {\rm gr ~ cm^{-3}}.
  \label{eq:density}
\end{eqnarray}
It is seen that the number density in the breakout zone lies in the range $10^{11} - 10^{14}$ cm$^{-3}$ for anticipated conditions.
Note the scaling $v_b\propto \rho_b^{0.51}E_{exp}^{0.9}M_*^{-0.64}$, $E_s\propto \rho_b^{-1.65}E_{exp}^{-0.15}M_*^{0.1}$.

\begin{figure}
\begin{center}
\includegraphics[width=8.2cm,keepaspectratio]{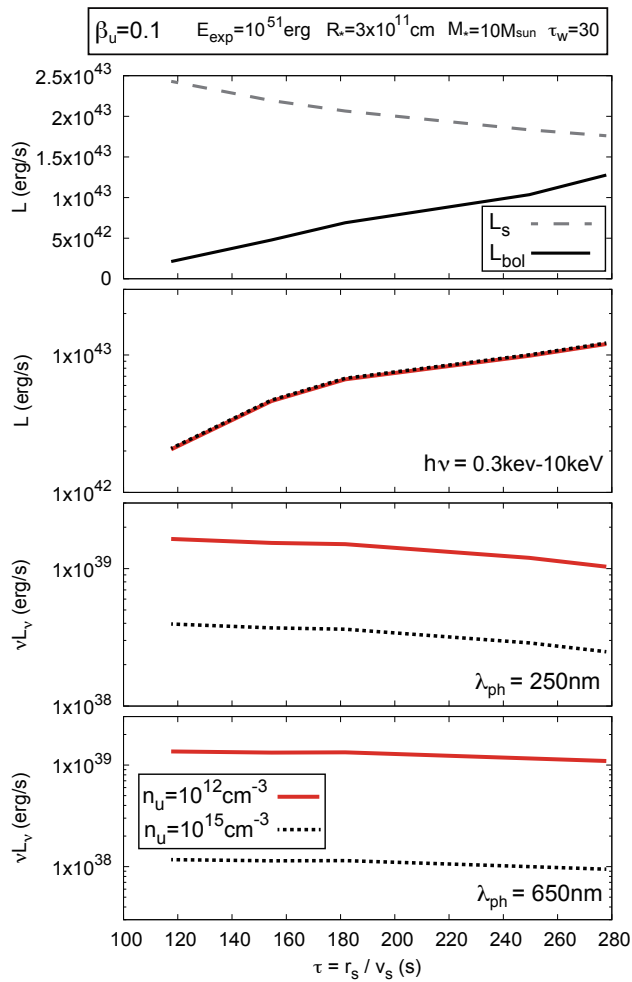} 
\end{center}
\caption{Estimated light curves based on the simulation results of $\beta_{\rm u} = 0.1$. The assumed model parameters for the shock dynamics are $E_{exp}= 10^{51}~{\rm erg}$, $R_* = 3\times 10^{11}~{\rm cm}$, $M_* = 10M_{\odot}$ and $\tau_w = 30$.
  In the top panel, the black solid line shows the bolometric luminosity.
  In addition, the shock luminosity is also dispayed with grey dashed line for comparison.
 The lower panels display the luminosities at a given band: 
   X-ray  ($h\nu =  0.3 - 10~{\rm keV}$), NUV ($\lambda_{\rm ph} = 250~{\rm nm}$) and optical ($\lambda_{\rm ph} = 650~{\rm nm}$).  
  The dashed black lines are the results computed from the fiducial simulations which assume $n_{\rm u} = 10^{15}~{\rm cm}^{-3}$.
  The solid red lines are the light curves duduced for $n_{\rm u} = 10^{12}~{\rm cm}^{-3}$.
  Note that the bolometric light curve does not vary with the density $n_{\rm u}$ since the shock structure does not change.
}
\label{LCB1}
\end{figure}

\begin{figure}
\begin{center}
\includegraphics[width=8.2cm,keepaspectratio]{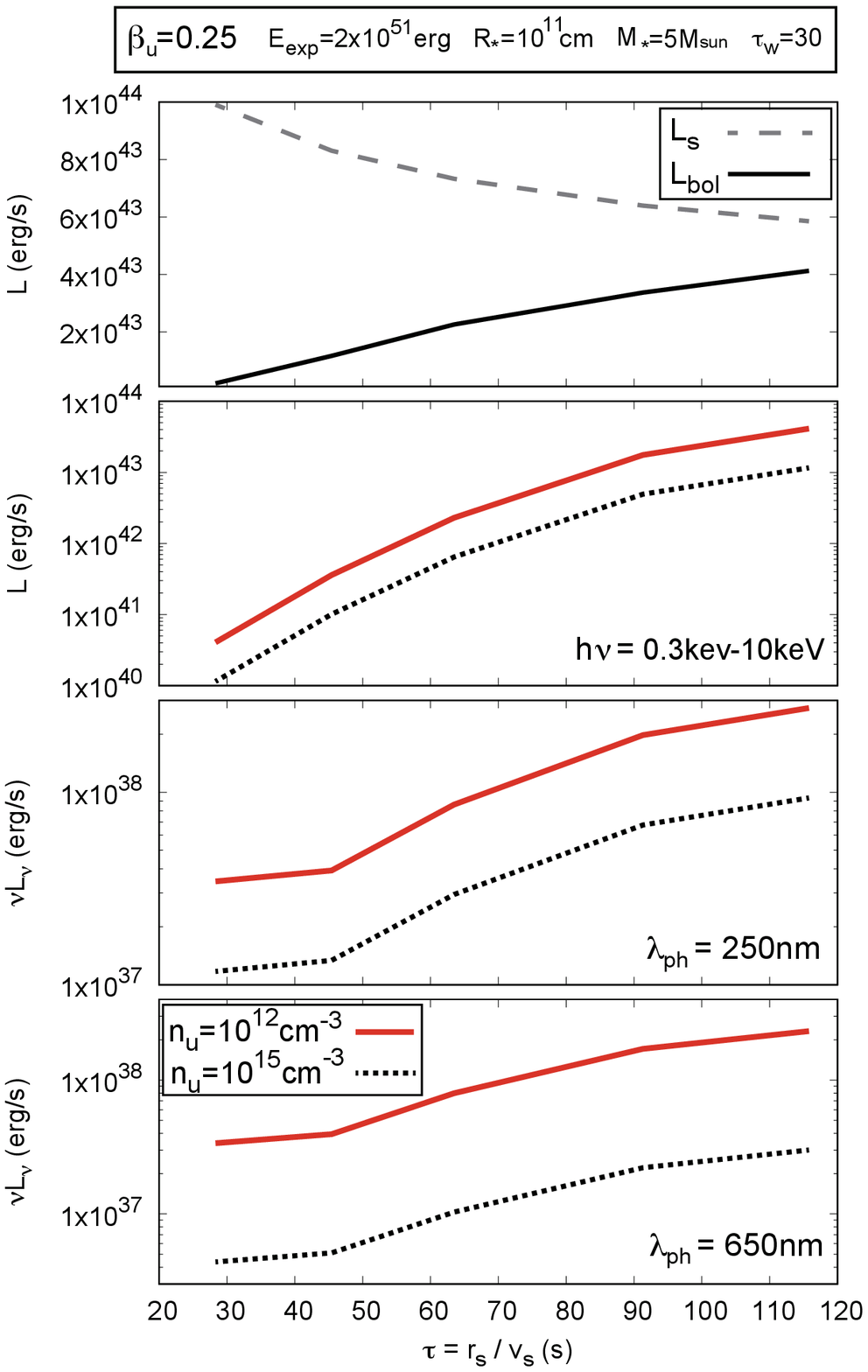} 
\end{center}
\caption{Same as Fig. \ref{LCB1}, but for $\beta_{\rm u} = 0.25$.  The assumed model parameters for the shock dynamics are $E_{exp}=2\times 10^{51}~{\rm erg}$, $R_* =  10^{11}~{\rm cm}$, $M_* = 5M_{\odot}$ and $\tau_w = 30$.
}
\label{LCB25}
\end{figure}

\begin{figure}
\begin{center}
\includegraphics[width=8.2cm,keepaspectratio]{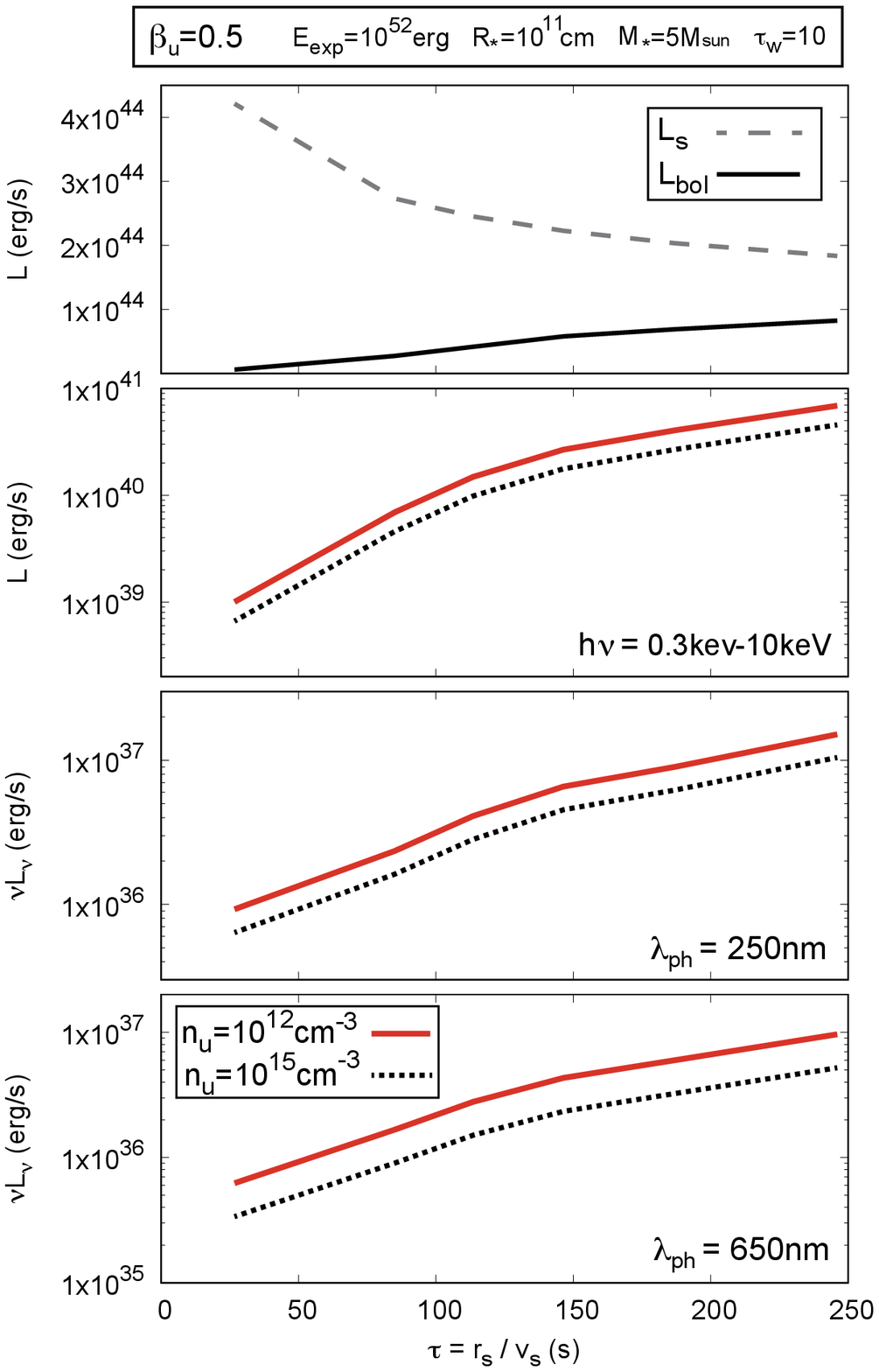} 
\end{center}
\caption{Same as Fig. \ref{LCB1}, but for $\beta_{\rm u} = 0.5$.
 The assumed model parameters for the shock dynamics are $E_{exp}= 10^{52}~{\rm erg}$, $R_* =  10^{11}~{\rm cm}$, $M_* = 5M_{\odot}$ and $\tau_w = 10$.
}
\label{LCB5}
\end{figure}

We now use the above results in conjunction with the simulations to compute lightcurves in different bands. 
We adopt the following procedure: First, we ignore, for simplicity, the dependence of the shock velocity on $\tau_s$ 
and take it to be constant during the breakout phase, which is justified by virtue of the very weak dependence in Eq. (\ref{eq:v0}).
For each RMS case simulated we choose a set of values for $E_{exp}$, $M_*$, $R_*$ and $\tau_w$, for which $v_s$ in
Eq. (\ref{eq:v0}) equals the simulated shock velocity (i.e., $\beta_{\rm u}=v_s / c = 0.1$, $0.25$ or $0.5$), as indicated in the titles
of figures \ref{LCB1}-\ref{LCB5}.   Next, we identify $\tau_s$ with the shock width, specifically, the
pair-unloaded optical depth of the shock, $\tau_s = \int_0^{l_s} n \sigma_T dx$,  measured in the simulations
from the upstream boundary at $\tau=0$ to the downstream point $l_s$ where $\beta=\beta_{\rm u}/6.5$.  For
each value of $\tau_s$ we then obtain the escape fraction $f_{esc}$ from the analysis in section \ref{Sec:structure} (see Figs. \ref{vsub} and \ref{vB5}),
and the shock energy and expansion time from Eqs. (\ref{eq:SW1}) and (\ref{eq:t}).
The bolometric luminosity of the breakout emission at time $t$ is then given by $L_{bol} = f_{esc} L_s$, 
where $L_s = E_s / t$ denotes the mean change in shock energy with time.
The spectral luminosity is given in terms of the flux density $f_\nu$ shown in Fig. (\ref{SPDS}) as $L_\nu = f_\nu (L_s/F_b)$.
Note that $\int_0^\infty f_\nu d\nu =f_{esc} F_b$, so that $\int_0^\infty L_\nu d\nu =f_{esc} L_s=L_{bol}$, as required.

We note that for a given velocity $\beta_{\rm u}=v_b/c$, once a value of the product $\tau_w R_*$ is chosen the breakout density 
is fixed by Eq. (\ref{eq:density}).  This means that the upstream density invoked in our simulations is inconsistent with the breakout
density.   While the bolometric luminosity is independent of the density, the spectral luminosity below the peak does depend on it. 
Performing additional simulations with different densities is highly demanding. As a compromise we exhibit below  lightcurves 
computed for the densities $n_{\rm u}=10^{15}$ cm$^{-3}$ and $10^{12}$ cm$^{-3}$ employed in the simulations presented in the preceding sections.  
For the choice of parameters in \ref{LCB1}-\ref{LCB5} Eq. (\ref{eq:density}) yields a breakout density of $3\times10^{13}$, 
  $1.5\times10^{13}$ and $1.2\times10^{11}$ cm$^{-3}$ for $\beta_{\rm u}=0.1, 0.25$ and $0.5$, respectively. Note that $f_{\pm} \sim 10$ is employed for $\beta_{\rm u}=0.5$ (see Fig. \ref{TnB5}).
In practice the density 
will decline during the gradual breakout by up to a factor of a few. 

In each of the figures  \ref{LCB1}-\ref{LCB5} we show 3 sets of light curves
computed for the choice of model parameters indicated in each figure title, which
are suitable for Wolf-Rayet stars surrounded by optically thick wind, that upon exploding release energy in the range
$E_{exp} = 10^{51}-10^{52}~{\rm erg}$.
From top to bottom, the panels in each figures show
  the  bolometric, X-ray ($h\nu = 0.3 - 10~{\rm keV}$), NUV ($\lambda_{\rm ph} = 250~{\rm nm}$) and optical ($\lambda_{\rm ph} = 650~{\rm nm}$) light curves,  where $\lambda_{\rm ph}$ is the wavelength of photons.
%
%
 In addition to the light curves produced based on the fiducial simulations ($n_{\rm u} = 10^{15}~{\rm cm}^{-3}$), we also plot the estimates 
for $n_{\rm u} = 10^{12}~{\rm cm}^{-3}$. 
Here we assume that the flux ratio at a given energy does not vary largely with the escape fraction, and
the estimates are made by  multiplying the fiducial light curves by a constant factor determined by the flux ratio between the simulations for
$n_{\rm u} = 10^{15}~{\rm cm}^{-3}$ and $n_{\rm u} = 10^{12}~{\rm cm}^{-3}$ shown in Fig. \ref{SPndep}.  Given our estimates
of breakout density,
  realistic light curves are expected to lie between the two shown for $\beta_{\rm u}=0.1$ and  $\beta_{\rm u}=0.25$ but  closer to the upper 
  curve (for $n_{\rm u}=10^{12}$), while   $\beta_{\rm u}=0.5$ is expected to be slightly above the two curves.
Note that $L_s=E_s/t \propto E_{exp}^{1.5}M_*^{-1.1}(\tau_wR_*)^{0.1}$ depends very weakly on the wind's opacity. 
Substantially larger luminosities require larger explosion energies and smaller ejecta mass.



As seen in the figure, while the energy deposition rate $L_s$ declines with time, the emission becomes brighter as the escape fraction increases.
Since the escape fraction at the latest time is large ($f_{esc} \gtrsim 0.5$),
the light curves are expected to reach the peak soon after and connect to cooling envelope emission.
The bolometric luminosity represents the emission 
at X-ray and gamma-rays: $\sim {\rm keV}$ for $\beta_{\rm u} = 0.1$, $\sim 10~{\rm keV}$ for $\beta_{\rm u} = 0.25$
and $\sim 100~{\rm keV}$ for $\beta_{\rm u} = 0.5$.
Although the optical/UV are much dimmer, it is much brighter than
the naive expectation from Wien spectra as mentioned in the previous section.

\section{Comparison to SN 2008D/XRT 080109}
\label{sec:comparison}
The leading candidate of a SN shock breakout from a dense stellar wind is SN2008D/XRT 080109  \citep{S08,modjaz2009}. The X-ray flash seen in this SN has a rise time of $50-100$ s, followed by a shallow power-law decay that lasts for about 300 s. The peak luminosity is $\sim 4 \times 10^{43}$ erg/s and the time integrated spectrum over the entire observed emission, which is dominated by the slow decay phase, is consistent with a flat power-law, $\nu F_\nu \approx {\rm const}$. After $\sim 300$s the X-ray light curve drops sharply. 

A shock breakout through a thick wind is one of the leading models for this X-ray flash, due to its relatively long duration \citep{Chevalier08,balberg2011,svirski2014a,ILN19}.  According to this model the rising part of the light curve is produced during the shock breakout episode. The transition to a collisionless shock takes place near the peak and the shallow power-law originates from the propagation of the collisionless shock in the optically thick wind \citep{svirski2014a}. The sharp decline after $\sim 300$s marks the transition of the shock to the optically thin wind region. Previous studies examined this model based on the time and energy scales of the entire emission \citep{Chevalier08,balberg2011}, and based on a detailed comparison of the light curve and spectrum from the collisionless phase (post breakout) to a theoretical model \citep{svirski2014a}.  With our current study we can examine the compatibility of the predicted spectrum and the total luminosity in the XRT energy window (0.3-10 keV) with the observations.

 According to the modelling of the post-peak emission \citep{svirski2014a} we find a breakout velocity of $\beta_{\rm u} \approx 0.2-0.4$ and a breakout radius $R_{bo} \sim 10^{12}$ cm. Thus, the parameters adopted in figure \ref{LCB25}  ($\beta_{u} =0.25$, $n_{\rm u}=10^{12} {\rm~cm^{_3}}$) are in agreement with those inferred for XRT 080109. The luminosity predicted for this choice of parameters in the XRT spectral window is in general agreement with the  observations. The $\nu F_\nu$ spectrum, as shown in figures  7 \& 8, peaks at early time (small value of $f$) around $\sim 20$ keV and at late times (near the peak) around $\sim 5$ keV. The peak of the integrated spectrum during the rising phase for $\beta_{\rm u}=0.25$ and $n_{\rm u}=10^{12} {\rm~cm^{_3}}$ is $\sim 10$ keV.  For a breakout velocity of $\beta_{\rm u}=0.2$ the peak of the integrated spectrum is smaller, $\sim 5$ keV. This implies that if  XRT 080109 is a wind breakout then the rising phase is harder than the decay. This is compatible with the the analysis of \citet{S08} that find a significant spectral softening during the outburst. Moreover, while the spectrum during the decay phase is expected to be a power-law with $\nu F_\nu \approx {\rm const}$ \citep{Svirski14b}, the breakout spectrum near the peak of $\nu F_\nu$ is expected to deviate from a power-law. It may be possible to identify such deviation in a reanalysis of the rising phase of XRT 080109.

 \section{Detectability}
 \label{sec:Detect}
 The short duration of the shock breakout makes it very challenging for detection. One interesting property of  shock breakout from a stellar wind is that for the anticipated range of conditions the bolometric luminosity emitted during the breakout phase is predicted to lie within a narrow range. It is almost independent of the progenitor radius and the wind opacity ($L \propto (R_*\tau_w)^{0.1})$, and its dependence on the explosion energy and the inverse of the ejecta mass is roughly linear. We estimate that for any progenitor type and any mass-loss history (prior to the explosion) the breakout luminosity is expected to fall in the range $L_s \sim 10^{43}-10^{44}$ erg. 
 Different progenitors and explosion conditions may be distinguished by the overall duration and total energy  of the breakout pulse.

\noindent{\underline{$\gamma$-rays}}:  The spectral  range at which the breakout luminosity is released depends on its velocity. For $\beta_{\rm u} \approx 0.35-0.5$ most of the energy is released in soft $\gamma$-rays. The luminosity for these velocities is $\sim 10^{44}$ erg/s (figure \ref{LCB5}), which makes them extremely hard to detect with current gamma-ray detectors. Even the Swift BAT, which is the most sensitive detector of soft gamma-rays currently in operation, can detect such flares only up 
to a distance of $\sim 5$ Mpc. It is therefore not surprising that shock breakouts from regular type Ib/c supernovae have not been detected in gamma-rays thus far.\\
 
\noindent{\underline{X-rays}}: A bright signal in the spectral window of most X-ray detectors, 0.3-10 keV, is expected for breakout velocities $\beta_{\rm u} \approx 0.1-0.35$. The range of luminosities in X-rays at these velocities is $\sim 1-5 \times 10^{43}$ erg/s (figures \ref{LCB1} and \ref{LCB25}) and the duration is $\sim 50$s for canonical parameters, and longer for a very massive wind. The most promising instrument for detection of such signals is eROSITA. Its single-scan sensitivity 
 is $\sim 10^{-13} {\rm~erg ~s^{-1}~ cm^{-2}}$ in the 0.5-10 keV band, and it scans the sky, spending $\sim 40$ s on each location within its 0.833 deg$^2$ field of view \citep{eROSITA12}. Thus, it can detect a shock breakout from a thick wind at a velocity of  $\beta_{\rm u} \approx 0.1-0.35$ up to a redshift of $z \approx 0.25$, which corresponds to a volume of $\sim 5 {\rm~Gpc^3}$. Given that the rate of type Ib/c SNe is $\sim 2.5 \times 10^4 {\rm~Gpc^{-3}~yr^{-1}}$ \citep{Li11}, and assuming that a shock breakout from a thick wind is common in this type of SNe (as suggested by the serendipitous detection of SN 2008D), we predict that eROSITA will detect roughly  one SN shock breakout signal every year. \\
 
\noindent{\underline{UV/optical}}: While the predicted UV and optical signals are much brighter than previously predicted, it is still rather faint. The brightest signal, expected for $\beta_{\rm u}=0.1$ in these bands,  has a luminosity of $\sim 10^{39}$ erg/s (figure \ref{LCB1}), which corresponds to an absolute AB magnitude  $M \sim -9$. Given the rate of type Ib/c SNe, such signal is much too faint for detection by any of the current and near future optical/UV surveys.

\section{Summary and conclusions}
\label{Sec:conclusion}

We performed Monte-Carlo simulations of photon-starved RMS that incorporate the leakage of photons from the shock,
for shock velocities $\beta_{\rm u}=0.1, 0.25$ (fast Newtonian regime)  and $\beta_{\rm u}=0.5$ (mildly relativistic regime).
We combined the simulation results with a shock propagation model to compute the signal emitted  during a 
gradual shock breakout from a stellar wind.  
This is the first prediction of the breakout emission from a wind obtained from first principles calculations. 
The main conclusion is that the flux emitted at frequencies below the SED peak (particularly the optical/UV band)
is much higher (by orders of magnitude) from that hitherto anticipated by naively invoking a Wien spectrum downstream
of the shock.  A detailed summary of the main results follows:

(i)   We find that in the fast Newtonian RMS  ($\beta_{\rm u} = 0.1$ and $0.25$) the temperature in the immediate downstream
decreases with increasing radiative losses, in agreement with the prediction of the analytical model of \cite{ILN19}.
This results from the enhancement of the photon density with increasing shock compression ratio (due to a slower photon diffusion downstream).
As a consequence, the peak energy of the breakout emission ($E_p \sim 3 k T \sim 1$ keV for $\beta_{\rm u}=0.1$ and $\sim 10~{\rm keV}$  for $\beta_{\rm u}=0.25$)
shifts to lower values as the luminosity increases.
This  might give rise to a power-law feature in the time-integrated spectrum (as suggested in Fig. \ref{SPDS} for the $\beta_{\rm u}=0.25$ case) which could explain the 
spectrum of the shock breakout candidate XRT080109, as discussed in \citet{ILN19}.
In contrast, in mildly relativistic shocks ($\beta_{\rm u} = 0.5$) the temperature is regulated by pair creation and is, therefore,
quite insensitive to the escape fraction.
As a result, the peak energy of the emission is expected to be fixed during the breakout phase at around $E_p \sim 100~{\rm keV}$.
Our analysis predicts a detection rate of about one SN shock breakout event per year by eROSITA.

(ii) The time-resolved spectra of escaping photons 
are well described by a Wien spectrum at the energies around the SED peak and above.
However, there is substantial softening of the portion of the spectrum below the peak ($f_\nu \propto \nu^{0 - 0.5}$) which extends 
down to the frequency below which free-free absorption becomes important.
This implies that the soft emission should be much brighter than the naive expectation assuming a Wien spectrum 
by orders of magnitude.  
Although it is not clear at present, the subshock found for  $\beta_{\rm u} = 0.5$
may give rise to efficient particle acceleration.
In this case, the resulting spectrum may be affected by the accelerated pairs. Further investigation is necessary to pin down this issue. 

(iii) The computed light curves show a gradual rise over tens to hundreds of seconds, depending on parameters, in all bands, except 
for the optical lightcurve of the $\beta_{\rm u}=0.1$ shock which is flat.  The optical/UV luminosity is higher for slower shocks and lower 
breakout densities, as anticipated, and can reach $10^{40}$ ergs s$^{-1}$ for $\beta_{\rm u}=0.1$ at $250$ nm.  
Unfortunately, this is still too low to be detected by current and near future optical/UV surveys. 

(iv) The velocity profiles found for $\beta_{\rm u} = 0.1$
and $0.25$ are in good agreement with the analytical model of \citet{ILN19}.
This result confirms that the diffusion limit provides a reasonable approximation
for the radiation transfer at these velocities even when substantial energy is escaping from the shock.
%
 %
This is no longer true for the mildly relativistic shock $\beta_{\rm u} = 0.5$.
Contrary to the fast Newtonian shocks that exhibit a smooth shock profile for all 
escape fractions (at least up to $f_{esc}\sim 0.7$),  in the  $\beta_{\rm u} = 0.5$ case a 
subshock forms when the losses exceed a certain value (a few percents) 
and then continues to grow as $f_{esc}$ increases.  In practice, this subshock can accelerate 
the pairs to nonthermal energies which, in turn, may lead to a nonthermal high-energy 
spectral component.   Such a component was not included in our analysis. We plan to 
investigate the effect of the subshock on the evolution of the spectrum in a future work. 
%

 \section*{Acknowledgments}
 This work was supported by JSPS KAKENHI Grant Number JP19K03878, JP19H00693 and JP20H04751.
 Numerical computations and data analysis were carried out on
Hokusai BigWaterfall system at RIKEN,   XC50  at Center for Computational Astrophysics, National Astronomical Observatory of Japan and the Yukawa Institute Computer Facility.
This work was supported in part by a RIKEN Interdisciplinary Theoretical \&
Mathematical Science Program (iTHEMS) and a RIKEN pioneering project ``Evolution of Matter in the Universe (r-EMU)'' and
``Extreme precisions to Explore fundamental physics with Exotic particles (E3-Project)''.  
AL and EN acknowledge support by the Israel Science Foundation grant 1114/17. 
 

%

\section*{Data Availability}
The data underlying this article will be shared on reasonable request to the corresponding author.

\label{lastpage}

\end{document}